\documentclass[screen, authorversion]{acmart} 
\usepackage[T1]{fontenc}
\IfFileExists{libertine.sty}{}{\microtypesetup{expansion=false}}
\AtBeginDocument{%
  }

\setcopyright{none}

\acmSubmissionID{37}

\copyrightyear{2026}
\acmYear{2026}
\setcopyright{cc}
\setcctype{by}
\acmConference[SAP '26]{ACM Symposium on Applied Perception 2026}{August 26--27, 2026}{Rennes, France}
\acmBooktitle{ACM Symposium on Applied Perception 2026 (SAP '26), August 26--27, 2026, Rennes, France}
\acmDOI{10.1145/3821409.3833894}
\acmISBN{979-8-4007-2799-3/2026/08}

\begin{document}

\title{The Nocturnity Scale: Measuring the Sense of Being at Night in Virtual Urban Environments}

\author{Anthony Le Gourriérec}
\orcid{0009-0003-1115-3010}
\affiliation{
  \institution{Nantes Université, ENSA Nantes, École Centrale Nantes, CNRS, AAU-CRENAU, UMR 1563 }
  \institution{IMT Atlantique, Lab-STICC, UMR CNRS 6285}
  \institution{Observatoire de la nuit}
  \city{Nantes}
  \country{France}
}
\email{Anthony.Le-Gourrierec@ec-nantes.fr}

\author{Etienne Peillard}
\orcid{0000-0002-4429-670X}
\affiliation{
  \institution{IMT Atlantique, Lab-STICC, UMR CNRS 6285}
  \city{Brest}
  \country{France}
}
\email{etienne.peillard@imt-atlantique.fr}

\author{Nicolas Houel}
\orcid{0000-0003-2249-194X}
\affiliation{
    \institution{Nantes Université, ENSA Nantes, École Centrale Nantes, CNRS, AAU-CRENAU, UMR 1563}
    \institution{Observatoire de la nuit}
    \city{Nantes}
    \country{France}}
\email{nicolas.houel@lobservatoiredelanuit.fr}

\author{Myriam Servières}
\orcid{0000-0001-5749-1590}
\affiliation{
  \institution{Nantes Université, ENSA Nantes, École Centrale Nantes, CNRS, AAU-CRENAU, UMR 1563}
  \city{Nantes}
  \country{France}}
\email{myriam.servieres@ec-nantes.fr}
\renewcommand{\shortauthors}{Le Gourriérec, et al.}
\begin{abstract}
Nighttime environments are increasingly used in virtual urban studies, yet darkness alone does not fully recreate the subjective sense of being at night. Prior work suggests that this experience depends not only on the absence of daylight, but also on lighting structure, low-light perception, human activity, soundscape, and self-related states. However, no existing tool directly assesses this scene-dependent subjective experience.
This work introduces \textit{nocturnity} as the feeling of being at night elicited by a scene and proposes a theory-driven framework structured into three subscales: perceptual, activity, and inner-state nocturnity. Based on this framework, we develop a first candidate questionnaire for virtual urban environments. Developed through a literature-informed process and reviewed by two urban lighting experts, the scale comprises 42 Likert-type items, including three diagnostic subscales and complementary global and time-related items. This work provides a first operational basis for comparing virtual urban scenes according to their perceived nocturnity and supports future empirical validation.
\end{abstract}



\maketitle

\section{Introduction}
“Being at night” is not only about being in a dark environment. The experience of night in a virtual environment depends not only on darkness itself, but also on a broader set of cues, including changes in lighting characteristics due to artificial lighting \cite{onerDaytimeNighttimeImage2025}, variations in human activity such as traffic evolution \cite{romanowskaInvestigatingImpactWeather2022}, and shifts in subjective state, including perceived safety \cite{boomsmaFeelingSafeDark2014,lisParkLightingDark2024}

Quantifying such subjective experience would enable systematic comparisons across scenes and support the controlled study of individual and combined perceptual cues that are difficult to isolate in real environments. Yet, this perceptual dimension currently lacks a reliable and reproducible metric; to our knowledge, no existing tool directly addresses it.



We use the term ``nocturnity'' to designate the ``feeling of being at night.'' Measuring nocturnity would enable researchers and designers to compare alternative nighttime simulations, isolate the effects of specific cues, and assess whether a virtual environment conveys more than a visually darkened scene.

\section{Literature-based theoretical framework}
\begin{figure}[t]
    \centering
    \includegraphics[width=\linewidth]{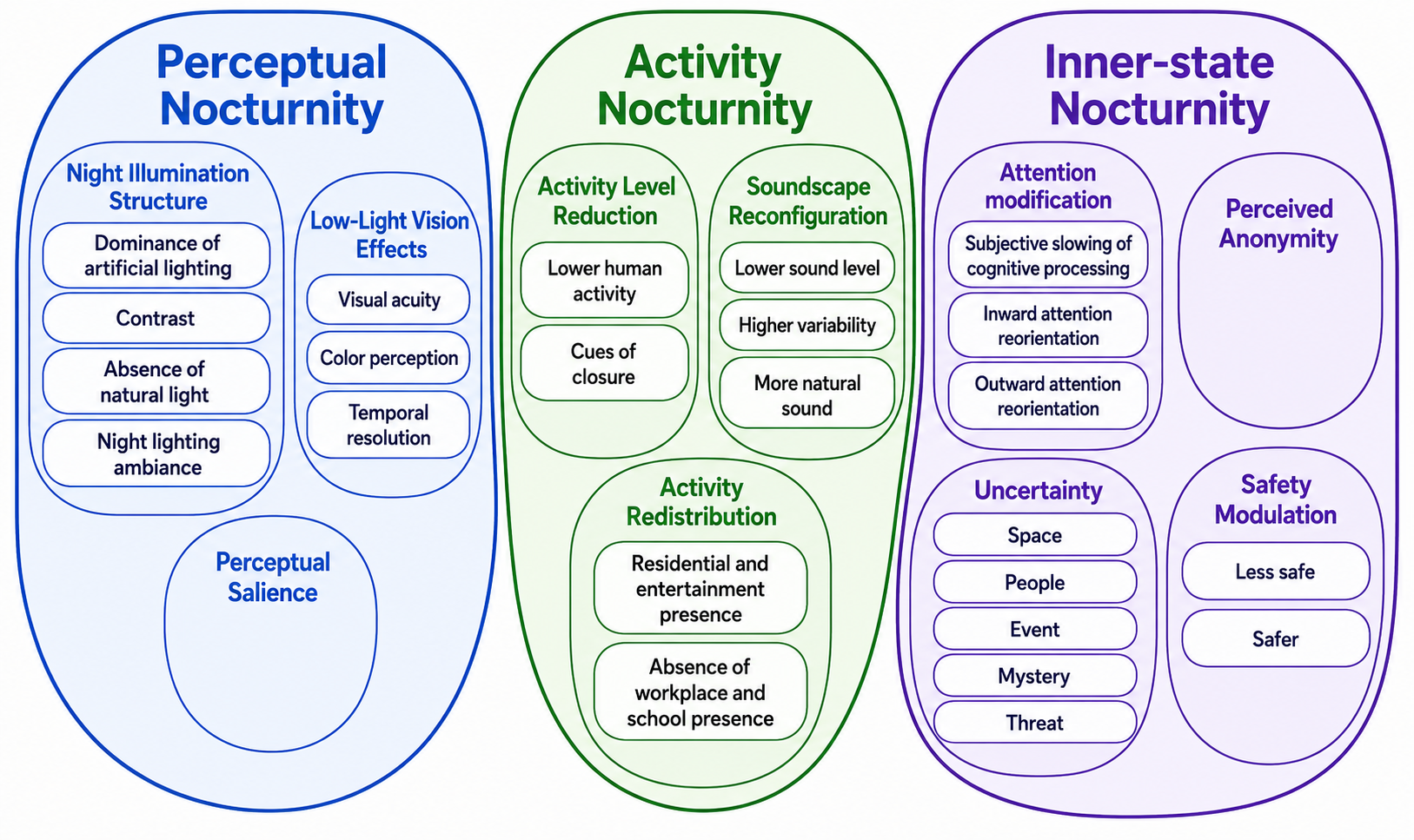}
    \caption{Venn-like hierarchical representation of the Nocturnity scale. Larger bubbles denote subscales, medium bubbles denote dimensions, and smaller bubbles denote facets.}
    \label{fig:nocturnity-venn}
    \Description{Venn-like hierarchical representation of the Nocturnity scale. Larger bubbles denote subscales, medium bubbles denote dimensions, and smaller bubbles denote facets.}
\end{figure}

We propose a literature-informed scale comprising 42 Likert-type items where \textit{nocturnity} is conceptualized as a hierarchical construct organized into three subscales: \textit{perceptual nocturnity}, \textit{activity nocturnity}, and \textit{inner-state nocturnity}–along with four additional global or time-related items. 
Each subscale is structured into dimensions and facets. The theory-driven framework, used for the generation of questionnaire items, is shown in Figure~\ref{fig:nocturnity-venn}.

\subsection{Definition of nocturnity subscales and dimensions}
The \textit{night illumination structure} dimension captures the night organization of lighting where artificial lighting becomes the dominant source of visibility~\cite{falchiNewWorldAtlas2016, onerDaytimeNighttimeImage2025}. Introduced within an otherwise dark environment, it produces strong spatial contrasts and selectively reveals elements of the scene while leaving others in darkness.  It is defined through several facets, including the \textit{dominance of artificial lighting}, the \textit{relative absence of natural light}, the resulting \textit{lighting contrast}, and the broader \textit{night lighting ambiance} emerging from the distribution, color, and intensity of light sources.
The \textit{low-light vision effects} dimension accounts perceptual transformation when the visual system transitions from photopic to mesopic and scotopic regimes \cite{zeleVisionMesopicScotopic2015, caoRodContributionsColor2008}.
 It is structured around facets corresponding to changes in \textit{visual acuity}, \textit{color perception}, and \textit{temporal resolution}, reflecting how the visual system adapts to low-light conditions.
The \textit{perceptual salience} dimension captures attention redistribution, where \textit{isolated elements}, \textit{subtle events}, and \textit{sparse stimuli} become more salient in nighttime than under daytime conditions \cite{morrisNightWalkingDarkness2011}.
These three dimensions define \textit{perceptual nocturnity}, which refers to the set of sensory cues through which an environment is perceived as night.

The \textit{activity level reduction} dimension captures the changes Human activity at night (lower traffic volumes, fewer pedestrians, and visible signs of closure \cite{romanowskaInvestigatingImpactWeather2022, duDailyRhythmUrban2024}), structured around facets such as \textit{lower human activity} and \textit{environmental indicators of closure}.
The \textit{soundscape reconfiguration} dimension reflects that nighttime soundscapes are not only quieter but also structurally different than daytime ones, shifting from continuous background noise to more intermittent and isolated auditory events \cite{
reygozaloAnalyzingNocturnalNoise2014}. Its facets include \textit{lower sound levels}, \textit{greater temporal variability}, and the \textit{emergence of natural sounds}.
The \textit{activity redistribution} dimension captures the nighttime activity redistribution across urban functions, concentrated in residential and leisure-oriented spaces, while work and education-related functions become less prominent \cite{duDailyRhythmUrban2024}, through facets such as \textit{residential and entertainment presence} and the \textit{absence of workplace and school-related activity}.
These dimensions define \textit{activity nocturnity}, which reflects the transformation of human activity patterns during the night.

The \textit{attention modification} dimension captures the attention reorientation linked to darkness (heightened vigilance or more introspective and reflective state \cite{morrisNightWalkingDarkness2011}), encompassing facets related to \textit{reduced alertness} as well as \textit{inward} and \textit{outward attentional reorientation}.
Darkness may induce a sense of illusory anonymity \cite{zhongGoodLampsAre2010, liuGenderModeratesEffect2018}.
The \textit{perceived anonymity} dimension reflects this subjective shift, capturing changes in perceived exposure and social awareness.
Darkness limits access to environmental information. As a result, individuals may experience increased ambiguity regarding space, others, and potential events \cite{toetEffectsPersonalRelevance2016, morrisNightWalkingDarkness2011}.
The \textit{uncertainty} dimension encompasses facets related to ambiguity in \textit{space}, \textit{people}, and \textit{events}, as well as its affective interpretations, ranging from \textit{threat} to \textit{mystery}.
The \textit{safety modulation} dimension captures the feelings of vulnerability or reassurance depending on the lighting context \cite{boomsmaFeelingSafeDark2014, lisParkLightingDark2024} 
through facets reflecting both \textit{increased vulnerability} and \textit{enhanced perceived safety}. 
Taken together, these dimensions define \textit{inner-state nocturnity}.

In addition to the three subscales, the questionnaire includes one global overall feeling of being at night item and three perceived time items that are not treated as a separate subscale.

\subsection{Expert interviews}
The candidate questionnaire was reviewed through semi-struct-ured one-on-one interviews with two experts in urban lighting. The interviews aimed to assess the relevance, clarity, and coverage of the candidate questionnaire.


Overall, the experts found the content of the questionnaire to be relevant and coherent with the targeted construct. 
Both experts requested clarification regarding items related to
perceptual salience, indicating that their intent was not immediately
clear.
One expert emphasized that some items may be context-dependent, particularly in socially active nighttime environments where perceived anonymity and reduced-activity cues may not apply in the same way. 
In addition, one item was
consistently judged as unclear and difficult to interpret, it was removed
from the candidate version.
Nevertheless,
the experts did not identify major omissions in the set of cues, supporting an initial level of content validity.

\section{Conclusion}
The present work is a first step toward a validated measure of nocturnity, defined here as the subjective feeling of ``being at night'' Based on the literature, we propose a theory-driven framework structured into three dimensions: \textit{perceptual nocturnity}, \textit{activity nocturnity}, and \textit{inner-state nocturnity}, and use it to develop a first candidate questionnaire. Expert feedback provided preliminary evidence of content validity and helped refine the scope and formulation of the instrument. Future work will evaluate the questionnaire in virtual reality, assessing its ability to discriminate between scenes with varying degrees of nocturnity, as well as its reliability and internal structure through standard psychometric analyses.
\bibliographystyle{ACM-Reference-Format}
\bibliography{bibtex}
\end{document}